\documentclass[aps,prl,twocolumn,showpacs,superscriptaddress,groupedaddress]{revtex4}  
\usepackage{graphicx}  
\usepackage{dcolumn}   
\usepackage{bm}        
\usepackage{amsthm}
\usepackage{amsmath}
\usepackage{amssymb}
\usepackage{amsfonts}
\usepackage{amscd}
\usepackage{ascmac}
\usepackage{enumerate}
\usepackage{bm}
\usepackage{latexsym}
\usepackage{color}

\newtheorem{thm} {Theorem}
\newtheorem{prp}[thm] {Proposition}


\def\;{{\hspace{0.3ex};\hspace{0.5ex}}}
\def\,{{\hspace{0,3ex},\hspace{0.5ex}}}
\def\({{\hspace{1.2ex}(}}

\def\R{{\mathbb{R}}}

\def\N{{\mathbb{N}}}

\def\C{{\mathbb{C}}}

\begin{document}

\title{Entanglement Concentration is Irreversible}

\author{Wataru Kumagai$^{1,2}$~~~Masahito  Hayashi$^{2,3}$\\
\textit{${}^1$Graduate School of Information Sciences, Tohoku University, Japan \\${}^2$Graduate School of Mathematics, Nagoya University, Japan,\\${}^3$Centre for Quantum Technologies, National University of Singapore, Singapore}}

\begin{abstract}

In quantum information theory, it is widely believed that entanglement concentration for bipartite pure states is asymptotically reversible.
In order to examine this, we give a precise formulation of the problem,
and show a trade-off relation between performance and reversibility, which implies the irreversibility of entanglement concentration. 
Then, we regard entanglement concentration as entangled state compression in an entanglement storage with lower dimension.
Because of the irreversibility of entanglement concentration, an initial state can not be completely recovered after the compression process and a loss inevitably arises in the process.
We numerically calculate this loss and also derive for it a highly accurate analytical approximation.

\end{abstract}

\pacs{03.65.Aa, 03.67.Bg}

\maketitle

\begin{figure}[ht]
 \hspace*{0em}\includegraphics[width=80mm, height=75mm,angle=0]{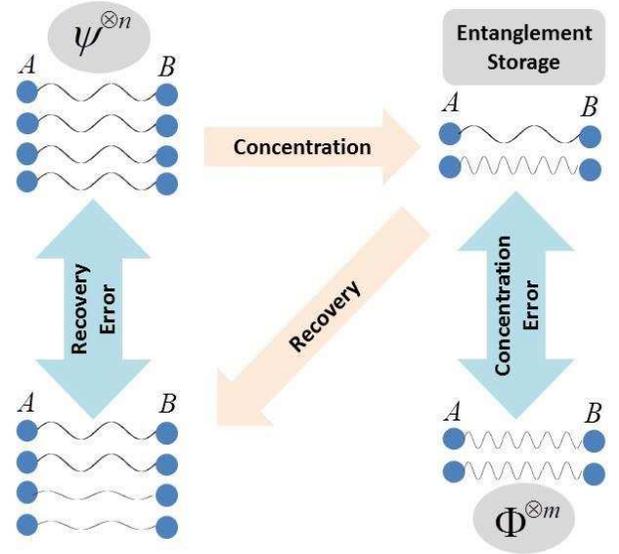}
 \caption{
The diagram of entanglement concentration and its recovery operation.
The initial (target) state consists of $n$ ($m$) copies of a pure (the EPR) state $\psi$ $(\Phi)$.
}
 \label{figure2}
\end{figure}

\hspace{-1.1em}{\it Irreversibility of Entanglement Concentration:}
Entanglement is an important resource in many quantum information processes,
and thus several types of conversion between entangled states have been studied in the literature.
One of the most typical such conversions is entanglement concentration, 
which approximately transforms multiple copies of a given pure state 
into multiple copies of the EPR state 
by using local operations and classical communication (LOCC).
Another fundamental conversion is entanglement dilution, which goes in the opposite direction, transforming  copies of the EPR state into copies of a target pure entangled state. 
The optimal rate of entanglement concentration is the von Neumann entropy 
of the partial density matrix of the initial state 
\cite{BBPS, HHT},
and equals the optimal rate of entanglement dilution. 
Therefore, entanglement concentration for bipartite pure states seems to be asymptotically reversible, as pointed out in  \cite{BPRST,AVC,JP,VC,YHHS}.
If entanglement concentration were reversible,  
we could use it asymptotically as lossless entanglement compression involving only LOCC operations.
That is, after compressing multiple copies of an entangled pure state in a lower dimensional storage system via entanglement concentration, the initial state could be recovered 
if the storage system had large enough number of copies of the EPR state with sufficient quality.
However, this kind of reversibility has not been sufficiently studied.
To examine whether asymptotically lossless entanglement compression can be realized,
this paper focuses on the process of concentration and its recovery operation, as illustrated in Fig. \ref{figure2},
where we transform by LOCC a number of copies of a bipartite pure state into  copies of an approximate EPR state of an entanglement storage system.
Afterwards, we apply LOCC to recover the initial state from that in the storage system.

For this process, we introduce two kinds of errors. 
Given a concentration operation $C$, 
the concentration error is defined as
\begin{eqnarray}\label{errorC}
e^{\cal C}_n(m,C|\psi)
:=1-F(C(\psi^{\otimes n}),\Phi^{\otimes m})^2,
\end{eqnarray}
where $F$ is the fidelity, $\Phi$ is the EPR state, and $m$ is the number of copies of the target EPR state after applying $C$.
From $(\ref{errorC})$, it is clear that a large concentration error represents low performance of the operation $C$.
The second error we need to introduce, the minimum recovery error, is
\begin{eqnarray}\label{errorR}
e^{\cal R}_n(C|\psi)
:=\min_{D:LOCC} 1-F(\psi^{\otimes n},D\circ C(\psi^{\otimes n}))^2,
\end{eqnarray}
where the minimization is taken over all LOCC recovery operations $D$.
As the minimum recovery error gets larger, 
the fidelity between the initial state and the { optimally} recovered state 
gets smaller.
In this sense, the minimum recovery error represents the degree of irreversibility of the concentration operation $C$.
Complete success of each operation gives zero error, whereas complete failure occurs when the error is 1. 
We note that the concentration error approaches $0$ if $n$ is sufficiently large compared to $m$ and a suitable concentration operation is taken. 
Similarly, the minimum recovery error is $0$ when the identity is chosen as the concentration operation.
Therefore, each error can individually attain the value zero. 
However, those errors are not compatible with each other, as shown below.

In order to clarify the trade-off between performance and reversibility of entanglement concentration,
we consider the minimal concentration-recovery error (MCRE) defined as
\begin{eqnarray}
\delta_n(\psi):=\min_{m\in\N,C:LOCC} \left\{e^{\cal C}_n(m,C|\psi)+e^{\cal R}_n(C|\psi)\right\}.
\label{MCRE}
\end{eqnarray}
We note that 
the same concentration operation $C$ is in both errors in the minimization of (\ref{MCRE});
i.e. the minimization over $C$ cannot be carried out independently for each error.
The MCRE obviously takes values between $0$ and $2$, and represents the degree of compatibility of the two operations.
In particular, if perfect reversible entanglement is possible in the asymptotic limit, 
the concentration and the minimum recovery errors simultaneously go to $0$,
and so does the MCRE.
However, we show below that $\delta_n(\psi)$ does not tend to $0$.
\begin{thm}\label{MCREineq}
 $\displaystyle\lim_{n\to\infty}\delta_n(\psi) =1$ for any bipartite pure entangled state $\psi$ that is not maximally entangled.
\end{thm}

Theorem \ref{MCREineq} shows a trade-off between the concentration error and the minimum recovery error, 
that is, the smaller one of the errors is, the larger the other error becomes under the constraint that their sum is $1$.
In particular, when the concentration error asymptotically goes to the minimum value $0$, the minimum recovery error always goes to the worst value $1$,
and hence, 
perfect entanglement concentration is completely irreversible.

~

\noindent{\it {Numerical Verification and Outline of the Proof:}}
To numerically demonstrate and analytically prove Theorem \ref{MCREineq},
we introduce the minimum transition error from a state $\psi$ to another state $\phi$
as 
\begin{eqnarray}
&d(\psi\to\phi)
:=\displaystyle\min_{E:LOCC}(1-F(E(\psi),\phi)^2).& \nonumber
\end{eqnarray}
As shown in Supplemental Material, 
the MCRE has the following representation. 

\begin{prp}\label{prpeq}
\begin{eqnarray}\label{eq}
\hspace{-0em}\delta_n(\psi)
= \min_{m\in\N} d(\psi^{\otimes n}\to\Phi^{\otimes {m}}) + d(\Phi^{\otimes {m}}\to\psi^{\otimes n}).
\end{eqnarray}
\end{prp}

Since $\Phi$ is the EPR state,
the first and second terms of the right hand side of (\ref{eq}) have operational meaning: 
the optimal errors of entanglement concentration and dilution for $\psi$ in non-asymptotic settings, respectively.

\begin{figure}[t]
 \hspace*{0em}\includegraphics[width=80mm, height=50mm,angle=0]{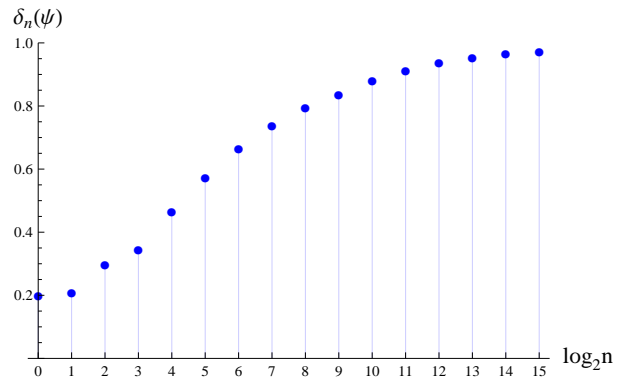}
 \caption{
Plot of $\delta_n(\psi)$ vs. ${\rm log}_2 n$, where $\psi$ is the pure entangled state $\sqrt{0.1}|00\rangle+\sqrt{0.9}|11\rangle$ of a two qubit system. 
We note that the MCRE
$\delta_n(\psi)$ approaches $1$ as $n$ goes to $\infty$.
}
 \label{Delta}
\end{figure}

Let us first focus on how to calculate the MCRE numerically and verify the validity of Theorem \ref{MCREineq}.
Here, we would like to point out that the MCRE was defined by minimizing 
the sum of the concentration error 
and the minimum recovery error 
over common possible LOCC operations, and hence its numerical calculation is highly non-trivial. 
Proposition \ref{prpeq} reduces it to 
the individual minimizations of entanglement concentration and dilution, and it enables us to numerically calculate the MCRE (see Supplemental Material).
The behavior of the MCRE $\delta_n(\psi)$ as a function of ${\rm log}_2 n$ is shown in Fig. \ref{Delta},
where we see that $\delta_n(\psi)$ converges to $1$ when $n$ goes to $\infty$, as stated in Theorem \ref{MCREineq}.

Next, we prove Theorem \ref{MCREineq} by analyzing the asymptotic behavior of the two kinds of errors in the right hand side of (\ref{eq}).
To investigate these errors, we proceed similarly as in the analysis of classical data processing in \cite{NH13,Hay06}, and expand the minimizer $m$ in (\ref{eq}) as $an+b\sqrt{n}$. 
We focus on the coefficients $a$ and $b$, called the first and the second order rates, respectively. 
The following proposition is essential to derive Theorem \ref{MCREineq} and, moreover, it provides a very useful asymptotic formula for the optimal entanglement concentration and dilution errors.

\begin{prp}\label{2-order}
The following holds for any bipartite pure entangled state $\psi$ that is not maximally entangled:
\begin{eqnarray}
&&{\lim_{n\to\infty}}d(\psi^{\otimes n} \to \Phi^{\otimes an+b\sqrt{n}})\nonumber\\
&=&1-{\lim_{n\to\infty}}d(\Phi^{\otimes an+b\sqrt{n}} \to \psi^{\otimes n})\nonumber\\
&=&\left\{
\begin{array}{ll}
0 & \mathrm{if}~a<S_{\psi} \\
G\left(\frac{b}{\sqrt{V_{\psi}}}\right) & \mathrm{if}~a=S_{\psi} \\
1 & \mathrm{if}~a>S_{\psi},
\end{array}
\right.
\end{eqnarray}
where $S_{\psi}$ is the von Neumann entropy of the partial state $\mathrm{Tr}_B\psi$, $V_{\psi}:=\mathrm{Tr}\left\{(\mathrm{Tr}_B\psi)(-\mathrm{log}(\mathrm{Tr}_B\psi)-S_{\psi}{\bf I}_A)^2\right\}$ and 
$G$ is the cumulative distribution function of the standard normal distribution.
\end{prp}

Propositions \ref{prpeq} and \ref{2-order} prove Theorem \ref{MCREineq}
because the sum of $d(\psi^{\otimes n} \to \Phi^{\otimes an+b\sqrt{n}})$ and $d(\Phi^{\otimes an+b\sqrt{n}} \to \psi^{\otimes n})$ goes to 1 as $n$ tends to $\infty$ for any $a$ and $b$.
In particular, when $a$ is the optimal rate $S_{\psi}$, Proposition \ref{2-order} gives a precise description of the asymptotic behavior of both errors for arbitrary $b$, which has not been given in the previous related literature \cite{BBPS,HKMMW,HW03,HL04}. 
It also states that the sum of limits of both errors is always $1$ for any $b$, as illustrated in Fig. \ref{boundary}. 
In Supplemental Material, we provide the detailed proofs of Proposition \ref{2-order} and Theorem \ref{MCREineq} as well as some remarks, including the relation with Hayden and Winter \cite{HW03} and Harrow and Lo \cite{HL04}.

\begin{figure}[t]
 \hspace*{0em}\includegraphics[width=80mm, height=50mm,angle=0]{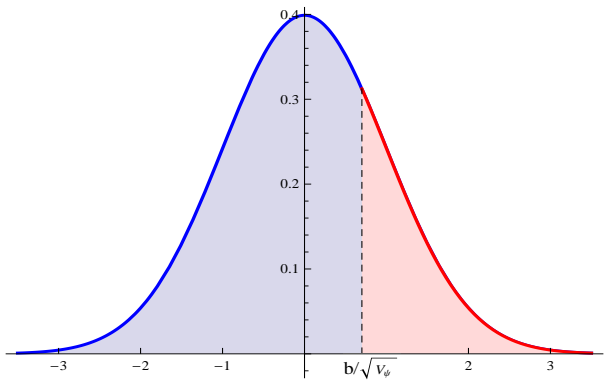}
 \caption{
The thick line is the normal distribution function.
The left blue area is $G({b}/{\sqrt{V_{\psi}}})$ and the right red area is 
 $1-G({b}/{\sqrt{V_{\psi}}})$, and they coincide with the limits of  $d(\psi^{\otimes n} \to \Phi^{\otimes S_{\psi}n+b\sqrt{n}})$ and $d(\Phi^{\otimes S_{\psi}n+b\sqrt{n}} \to \psi^{\otimes n})$, respectively.
These two quantities always sum up to $1$ for any second order rate $b$.
}
 \label{boundary}
\end{figure}

~

\noindent{\it {LOCC Compression  Process:}}
Here, we regard entanglement concentration as entangled state compression in an entanglement storage (see FIG. \ref{figure2}).
Note that, as stated above, the initial state can not be completely recovered after the concentration since this process is irreversible.
In order to realize the compatibility of entanglement concentration and its reversibility, we accept that the number $N$ of recovered copies of $\psi$ after concentration is smaller than the number $n$ of copies of the initial state.
Our aim is to compute how many copies $N$ of the initial
state can be recovered after concentration
assuming some error margin in the compression process.
In such a situation, the irreversibility of a concentration operation $C$ is conveniently represented by the generalized minimum recovery error
\begin{eqnarray}
e^{\cal R}_n(C,N|\psi)
:=\min_{D:LOCC}1-F(\psi^{\otimes N},D\circ C(\psi^{\otimes n}))^2,
\end{eqnarray}
and the error of the compression-recovery process is measured by the generalized MCRE defined as
\begin{eqnarray}
\delta_n(N|\psi)
:=\min_{m\in\N,C:LOCC}\{e^{\cal C}_n(m,C|\psi)+e^{\cal R}_n(C,N|\psi)\}.
\label{MCRE2}
\end{eqnarray}
In particular, the previous MCRE $\delta_n(\psi)$ in (\ref{MCRE}) coincides with $\delta_n(n|\psi)$.
Our purpose is to maximize the number of recovered copies after compression with an error margin $\epsilon$,
and thus, we consider the following quantity
\begin{eqnarray}
N_n(\epsilon|\psi):=\max\left\{N| \delta_n(N|\psi)\le\epsilon \right\}.
\label{Nn}
\end{eqnarray}
Hence, we lose at least $n-N_n(\epsilon|\psi)$ copies of $\psi$
when we restrict the error of entanglement compression. 
For a practical use of entanglement compression, 
it is important to evaluate how large the minimum loss $n-N_n(\epsilon|\psi)$ is.

\begin{figure}[t]
 \hspace*{0em}\includegraphics[width=80mm, height=55mm,angle=0]{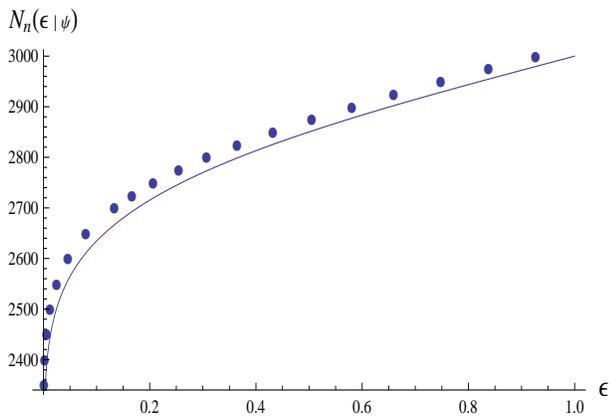}
 \caption{
Plot of $N_n(\epsilon|\psi)$ vs. $\epsilon$ for $\psi=\sqrt{0.1}|00\rangle+\sqrt{0.9}|11\rangle$ and $n=3000$.
The dots are numerical values of $N_n(\epsilon|\psi)$.
The solid line is the graph of $n-{2\sqrt{V_{\psi}}}{S_{\psi}^{-1}}G^{-1}(1-{\epsilon}/{2})\sqrt{n}$. 
We see that the the asymptotic expansion in Theorem \ref{2nd.rate} provides a good approximation to $N_n(\epsilon|\psi)$ if $n$ is large enough.
}
 \label{fig.Nn}
\end{figure}

To compute $N_n(\epsilon|\psi)$, 
we first consider the generalized MCRE $\delta_n(N|\psi)$, which can be regarded as the inverse function of $N_n(\epsilon|\psi)$.
The following proposition is a generalization of Proposition \ref{prpeq}.
\begin{prp}\label{prpeq2}
for $N\le n$, one has
\begin{eqnarray}
\delta_n(N|\psi)=\displaystyle\min_{m\in\N} d(\psi^{\otimes n}\to\Phi^{\otimes {m}}) + d(\Phi^{\otimes {m}}\to\psi^{\otimes N}).
\label{eq2}
\end{eqnarray}
\end{prp}
Just as Proposition \ref{prpeq}, 
Proposition \ref{prpeq2} reduces the calculation of the generalized MCRE to the individual minimizations over two independent operations, and so, enables us to calculate the generalized MCRE and $N_n(\epsilon|\psi)$ numerically (see Supplemental Material).
The results for $N_n(\epsilon|\psi)$ are the dots in Fig. \ref{fig.Nn}. 
In particular, we can see that the minimum loss $n-N_n(\epsilon|\psi)$ after compression reaches $10\%$ of the initial number even if the permissible error is $\epsilon=0.1$.
Since the rate of the loss is not so small and cannot be ignored; we need to evaluate it for a real implementation.
Then, as shown in Supplemental Material, the following approximation formula for $N_n(\epsilon|\psi)$ holds. 
\begin{thm}\label{2nd.rate}
Under a permissible error $0<\epsilon<1$ and up to order smaller than $\sqrt{n}$, one has the asymptotic expansion
\begin{eqnarray}
N_n(\epsilon|\psi)
=n-\frac{2\sqrt{V_{\psi}}}{S_{\psi}}G^{-1}\left(1-\frac{\epsilon}{2}\right)\sqrt{n}+o(\sqrt{n}).\nonumber
\end{eqnarray}
\end{thm}
In Fig. \ref{fig.Nn}, the solid line is the approximated formula of Theorem \ref{2nd.rate} as a function of $\epsilon$.
We can see that indeed the line approximates $N_n(\epsilon|\psi)$ numerically.

\begin{figure}[t]
 \hspace*{0em}\includegraphics[width=80mm, height=55mm,angle=0]{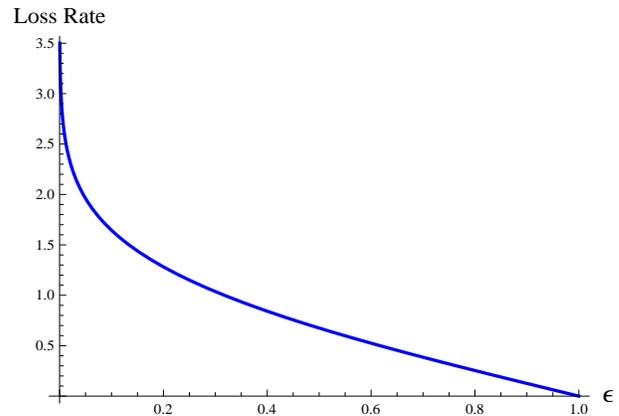}
 \caption{
The behavior of ${2\sqrt{V_{\psi}}}/{S_{\psi}}G^{-1}\left(1-\frac{\epsilon} {2}\right)$, i.e. the coefficient of $\sqrt{n}$ in (\ref{c-loss}), with respect to $\epsilon$  when $2\sqrt{V_{\psi}}S_{\psi}^{-1}=1$.
The function dramatically increases as $\epsilon$ gets smaller.
In particular, it approaches to $\infty$ and $0$ as the permissible error $\epsilon$ goes to $0$ and $1$, respectively. 
}
 \label{2rate2}
\end{figure}

When $n$-copies of an initial state are given, the minimum loss after compression is asymptotically evaluated from Theorem \ref{2nd.rate} to be
\begin{eqnarray}\label{c-loss}
n-N_n(\epsilon|\psi)
\cong\frac{2\sqrt{V_{\psi}}}{S_{\psi}}G^{-1}\left(1-\frac{\epsilon}{2}\right)\sqrt{n}.
\end{eqnarray}
The coefficient of $\sqrt{n}$ in (\ref{c-loss}) rapidly increases as $\epsilon$ gets smaller (see Fig. \ref{2rate2}) and, in particular, diverges to $\infty$ at $\epsilon=0$. 
Therefore, unlike the case when $0<\epsilon<1$, the loss $n-N_n(0|\psi)$ increases faster than order $\sqrt{n}$, and, 
the minimum loss after compression may not be ignored if $\epsilon$ is very small.
However, the minimum loss is order $\sqrt{n}$ as long as we accept a slight error, and the ratio of the loss over the initial number $n$ of copies gets smaller as $n$ gets larger.
Therefore, 
for a given permissible error $\epsilon$ in the compression operation,
if n is large enough,
 entanglement concentration works as the compression operation for an entangled pure state with the slight loss shown in (\ref{c-loss}).

~

\noindent{\it {Summary and Conclusion:}}~
In this letter, we have addressed entanglement concentration and its recovery operation for bipartite pure states. 
We have introduced the MCRE to simultaneously evaluate two corresponding errors in the process and derived the asymptotic  trade-off relation between them in Theorem \ref{MCREineq}.
As a consequence, 
it turns out that entanglement concentration is not reversible even in the asymptotic limit.
We have analyzed a compression process that consists of entanglement concentration and its recovery operation.
Due to the irreversibility of entanglement concentration,
some of the initial copies cannot be recovered 
when high quality entanglement concentration is required.
We have quantified the loss and obtained its analytical approximation in (\ref{c-loss}).
We have also derived Propositions \ref{prpeq} and \ref{prpeq2}, which enable us to numerically calculate the (generalized) MCRE and the maximum number 
of recoverable copies in the compression process.

\vspace{2em}

\hspace{-1.48em}{\it { Acknowledgment:}}~
The authors would like to thank Professor Emilio Bagan for helpful comments.
 WK acknowledges support from Grant-in-Aid for JSPS Fellows No. 233283. MH is partially supported by a MEXT Grant-in-Aid for Scientific Research (A) No. 23246071. 
The Center for Quantum Technologies is funded by the Singapore Ministry of Education and the National Research Foundation as part of the Research Centres of Excellence programme.


\newpage

\appendix

\section{Errors of entanglement concentration and dilution}\label{sec:Appendix}

\subsection{Remarks on Minimal Transition Errors}

For the sake of completeness, we introduce some related studies and give technical details of the minimum transition errors for entanglement concentration and dilution in this and the next subsection.
It has been already known that the minimum transition error $d(\psi^{\otimes n}\to\Phi^{\otimes an+b\sqrt{n}})$ of entanglement concentration goes to $0$ if the first order rate $a$ is strictly less than $S_{\psi}$ by Bennett et al. \cite{BBPS}, and does to $1$ if the rate is strictly greater than $S_{\psi}$ by Hayashi et al. \cite{HKMMW}.
Similarly, it has been also known that the minimum transition error $d(\Phi^{\otimes {an+b\sqrt{n}}}\to\psi^{\otimes n})$ of entanglement dilution goes to $0$ if the first order rate $a$ is strictly greater than $S_{\psi}$ by Bennett et al. \cite{BBPS}.
Therefore, the asymptotic behaviors of both minimum transition errors have been sufficiently analyzed unless $a$ is not $S_{\psi}$. 
On the other hand, 
when the first order rate $a$ strictly equals the optimal value $S_{\psi}$,
some existing studies suggest that the minimum transition errors $d(\psi^{\otimes n} \to \Phi^{\otimes S_{\psi}n+b\sqrt{n}})$ and $d(\Phi^{\otimes S_{\psi}n+b\sqrt{n}} \to \psi^{\otimes n})$ depend on the second order rate $b$.
As related studies about the second order asymptotics, 
we introduce Hayden and Winter \cite{HW03} and Harrow and Lo \cite{HL04} in the following. 
%
%
\subsection{Relation with Hayden and Winter \cite{HW03} and Harrow and Lo \cite{HL04}}

Here, we refer the relation with Hayden and Winter \cite{HW03}.
They focus on the evaluation for the classical communication cost of entanglement dilution, 
and in order to treat its asymptotic behavior, 
 show the importance of the second order rate in entanglement dilution.
They derive that a lower bound on the classical communication cost of a dilution protocol is given by the form of $b\sqrt{n}$ with some constant $b$ under a fidelity constraint.
They essentially evaluate 
the sum of smaller eigenvalues than $2^{-S_{\psi}n-b\sqrt{n}}$ of a partial state ${\rm Tr}_B\psi$, which appear as $K(b|\psi)=K(b,0|\psi)$ in the proof of Proposition 3, by using the central limit theorem in a part of the discussion. 


Next, we refer the relation with Harrow and Lo \cite{HL04}.
They show that, 
when $n$-copes of a pure entangled state $\psi$ can be approximately generated from $m$-copies of the EPR state $\Phi$ by LOCC under an error constraint, 
$m$ is bounded below by $S_{\psi}n+b\sqrt{n}$ with some constant $b$.
In addition, they represent the trade-off relation between the classical communication cost and success probability for a dilution protocol by using the second order rate $b$. 
Then, they essentially give the evaluation to $K(b|\psi)$
 by the Berry-Esse\'{e}n theorem.


From the analysis of \cite{HL04} and \cite{HW03}, we can see that $K(b|\psi)$ is described by the cumulative distribution function $G$ of the standard normal distribution as shown in (\ref{K}) .
However, they do not clarify the relation between $K(b|\psi)$ and the minimum transition errors $d(\psi^{\otimes n}\to\Phi^{\otimes an+b\sqrt{n}})$ and $d(\Phi^{\otimes an+b\sqrt{n}} \to \psi^{\otimes n})$.
On the other hand, we explicitly represent the relation between $K(b|\psi)$ and the minimum transition errors in the inequalities (\ref{r-1}), (\ref{r-2}), (\ref{s-1}) and (\ref{s-2}) in Proof of Proposition 3.

\subsection{Remarks on Incompatibility between Concentration and Recovery Errors}
We give an additional remark on the irreversibility of entanglement concentration.
Due to Theorem 1, 
both a concentration operation and its recovery operation can not be accurately performed. 
That is, there does not exist concentration operations satisfying both 
\begin{eqnarray}
&\displaystyle\lim_{n\to\infty} e^{\cal C}_n(m_n,C_n|\psi)=0,&\label{error.condition1}\\
&\displaystyle\lim_{n\to\infty} e^{\cal R}_n(C'_n|\psi)=0,&\label{error.condition2}
\end{eqnarray}
although there exist concentration operations $C_n:\mathcal{S}(\mathcal{H}_{AB}^{\otimes n})\to\mathcal{S}(\mathcal{H}_{2}^{\otimes m_n})$  satisfying (\ref{error.condition1})  and $C'_n:\mathcal{S}(\mathcal{H}_{AB}^{\otimes n})\to\mathcal{S}(\mathcal{H}_{2}^{\otimes m'_n})$ 
satisfying (\ref{error.condition2})
with the common first order rates $\lim m_n/n=\lim m'_n/n=S_{\psi}$.
The fact may look strange, however, can be comprehended by the argument of the second order rates. 
That is, 
when we expand $m_n$ and $m'_n$ as $S_{\psi}n+b\sqrt{n}+o(\sqrt{n})$ and $S_{\psi}n+b'\sqrt{n}+o(\sqrt{n})$, respectively, their second order rates $b$ and $b'$ are different.

~

\section{Numerical Calculation Algorithm}\label{sec:AppendixB}
\subsection{Calculation Algorithm for MCRE}
In order to obtain the MCRE $\delta_n(\psi)$ and the generalized MCRE $\delta_n(N|\psi)$, 
one has to minimize over all LOCC operations and ways to lead those values are non-trivial.
In the following, we provide algorithms to numerically calculate those values by using Propositions 2 and 4.

Since the original MCRE $\delta_n(\psi)$ equals the generalized MCRE $\delta_n(N|\psi)$ when $N=n$, we firstly state the algorithm for the generalized MCRE and then apply it to the original MCRE.
When we invoke (\ref{Ceq}) and (\ref{Deq}), Proposition 4 leads the following form which is described by the Schmidt coefficients of $\psi^{\otimes n}$,
\begin{eqnarray}
\delta_n(N|\psi)
&=&\min_{m\in\N}
\Big\{\sqrt{\frac{1}{2^{m}}} \displaystyle\sum_{i=1}^{J_{\psi^{\otimes n},2^{m}}}\sqrt{p_{\psi^{\otimes n},i}^{\downarrow}} \nonumber\\
&&+ \sqrt{\left(1-\frac{J_{\psi^{\otimes n},2^{m}}}{2^{m}}\right){\sum_{j=J_{\psi^{\otimes n},2^{m}}+1}^{2^m} p_{\psi^{\otimes n},j}^{\downarrow}}}\nonumber\\
&&+\sqrt{\sum_{i=1}^{2^{m}}p_{\psi^{\otimes N},i}^{\downarrow}}~\Big\}.
\end{eqnarray}

Let $\psi$ be $\sqrt{p}|00\rangle+\sqrt{1-p}|11\rangle$ with $0<p<1/2$. 
When $i$ satisfies 
\begin{eqnarray}
\sum_{k=0}^l\binom{n}{k}\le i<\sum_{k=0}^{l+1}\binom{n}{k},
\end{eqnarray}
 we have 
\begin{eqnarray}
p_{\psi^{\otimes n},i}^{\downarrow}
=p^l(1-p)^{n-l}
\end{eqnarray}
Then, $J_{\psi^{\otimes n},2^{m}}$ has the form of $\sum_{k=0}^l\binom{n}{k}$ and the $l$ is the maximal number under the condition that
\begin{eqnarray}
\frac{1-\sum_{k=0}^l\binom{n}{k}p^l(1-p)^{n-l}}{2^m-\sum_{k=0}^l\binom{n}{k}}
<p^{l-1}(1-p)^{n-l+1}.
\label{l}
\end{eqnarray}
Moreover, we have
\begin{eqnarray}
\sum_{i=1}^{2^{m}}p_{\psi^{\otimes N},i}^{\downarrow}
&=&\sum_{k=0}^{l'}\binom{N}{k}p^k(1-p)^{N-k}\label{long2}\\
&&+\left(2^m-\sum_{k=0}^{l'}\binom{N}{k}\right)p^{l'}(1-p)^{N-l'+1},
\nonumber
\end{eqnarray}
where $l'$ is the maximal number which satisfies 
\begin{eqnarray}
\sum_{k=0}^{l'}\binom{N}{k}\le 2^m.
\label{l'}
\end{eqnarray}
Thus, we obtain the following equation
\begin{eqnarray}
&&\delta_n(N|\psi)\nonumber\\
&=&\min_{m\in\N}
\Big[\sqrt{\frac{1}{2^{m}}} \displaystyle\sum_{k=0}^{l}\binom{n}{k}\sqrt{p^k(1-p)^{n-k}} 
\nonumber\\
&&
+ \sqrt{\left(1-\frac{\sum_{k=0}^l\binom{n}{k}}{2^{m}}\right) \left(1-\sum_{k=0}^{l}\binom{n}{k}p^k(1-p)^{n-k}\right)}\nonumber\\
&&
+\Big\{\sum_{k=0}^{l'}\binom{N}{k}p^k(1-p)^{N-k}\nonumber\\
&&\hspace{1em}+\Big(2^m-\sum_{k=0}^{l'}\binom{N}{k}\Big)p^{l'}(1-p)^{N-l'+1}\Big\}^{\frac{1}{2}}~\Big],
\label{long}
\end{eqnarray}
where $l$ is the maximal number which satisfies (\ref{l}) and $l'$ is the maximal number which satisfies (\ref{l'}).
Therefore, when $n$ and $N$ are concretely given, the right side in (\ref{long}) without $\min_{m\in\N}$ can be calculated for fixed $m$.
When the Schmidt rank of $\psi$ is $r(\psi)$,
$\min_{m\in\N}$ in (\ref{long}) can be replaced by $\min_{1\le m\le N\lceil \log_2 r(\psi)\rceil}$ because of the remark after Proofs of Propositions 2 and 4. 
Hence, $\delta_n(N|\psi)$ can be calculated by minimizing it with respect to $m$.
By summarizing the above discussion, the procedure of calculation of $\delta_n(N|\psi)$ for concrete $n$ and $N$ is described as follows:

\begin{description}
 \item[(i)]Fix $m=1,..., N\lceil \log_2 r(\psi)\rceil$.
 \item[(ii)]Maximize $l$ under the condition (\ref{l}).
 \item[(iii)]Maximize $l'$ under the condition (\ref{l'}).
 \item[(iv)]Calculate the right side of (\ref{long}) without $\min_{1\le m\le N}$ from $m$, $l$ and $l'$, and set it $\delta_n^{(m)}(N|\psi)$.
 \item[(v)]Repeat (i)-(iv) from $m=1$ to $m=N\lceil \log_2 r(\psi)\rceil$.
 \item[(vi)]Minimize $\delta_n^{(m)}(N|\psi)$ with respect to $m$.
\end{description}  
Since
\begin{eqnarray}
\delta_n(N|\psi)=\min_{1\le m\le N\lceil \log_2 r(\psi)\rceil}\delta_n^{(m)}(N|\psi),
\end{eqnarray}
we can calculate $\delta_n(N|\psi)$ from the above procedure.

We have derived the computable form for $\delta_n(N|\psi)$ from Proposition 4.
Similarly, Proposition 2 leads a computable form of the original MCRE as
\begin{eqnarray}
&&\delta_n(\psi)\nonumber\\
&=&\min_{1\le m\le n}
\Big[\sqrt{\frac{1}{2^{m}}} \displaystyle\sum_{k=0}^{l}\binom{n}{k}\sqrt{p^k(1-p)^{n-k}} 
\nonumber\\
&&
+ \sqrt{\left(1-\frac{\sum_{k=0}^l\binom{n}{k}}{2^{m}}\right) \left(1-\sum_{k=0}^{l}\binom{n}{k}p^k(1-p)^{n-k}\right)}\nonumber\\
&&
+\Big\{\sum_{k=0}^{l'}\binom{n}{k}p^k(1-p)^{n-k}\nonumber\\
&&\hspace{1em}+\Big(2^m-\sum_{k=0}^{l'}\binom{n}{k}\Big)p^{l'}(1-p)^{n-l'+1}\Big\}^{\frac{1}{2}}~\Big],
\label{long3}
\end{eqnarray}
where $l$ is the maximal number which satisfies (\ref{l}) and $l'$ is the maximal number which satisfies (\ref{l'}) with $N=n$.
The calculation procedure is described by (i)-(vi) with $N=n$. 
Then we can calculate values of $\delta_n(\psi)$ as shown FIG. 2 which corresponds to the case when $p=0.1$ in (\ref{long3}).

~

\subsection{Calculation of $N_n(\epsilon|\psi)$}
We have explained the calculation method for $\delta_n(N|\psi)$.
Note that $\delta_n(N|\psi)$ strictly increase with respect to $N$.
Then, a map from $N$ to $\epsilon_N:=\delta_n(N|\psi)$ is the inverse function of
$N_n(\cdot|\psi)$, that is,
$N_n(\epsilon_N|\psi)=N$ holds.
By using the equation, we can dot points $(\epsilon_N,N)$ as in FIG. 4.
Note that we firstly choose $N$, and then $\epsilon_N$ is determined by $N$.

~

\section{Proofs of Propositions and Theorems}

\subsection{ Proofs of Propositions 2 and 4}

Because Proposition 2 coincides with Proposition 4 when $N=n$,
it is enough to show Proposition 4.
For an arbitrary pure state $\psi\in\mathcal{H}_{AB}$, we denote the squared Schmidt coefficients of $\psi$ by $p_{\psi}=(p_{\psi,1},\cdot\cdot\cdot,p_{\psi,M})$. 
Let $\Phi_{L}=\sum_{i=1}^L \sqrt{1/L}|i\rangle |i\rangle$ be a maximally entangled state with the size $L$ on $\mathcal{H}_{L}:=\C^L\otimes\C^L$. 
When $p^{\downarrow}$ shows the probability distribution which is sorted in decreasing order for the components of $p$, we define the pure state $\eta_{\psi,L}$ in $\mathcal{H}_{L}$ as 
\begin{eqnarray}
\eta_{\psi,L}= 
\displaystyle\sum_{i=1}^{J_{\psi,L}}\sqrt{p_{\psi,i}^{\downarrow}}|i\rangle |i\rangle 
+ \sqrt{\frac{\sum_{j=J_{\psi,L}+1}^M p_{\psi,j}^{\downarrow}}{L-J_{\psi,L}}}\displaystyle\sum_{i=J_{\psi,L}+1}^L|i\rangle |i\rangle\nonumber
\end{eqnarray}
by using
\begin{eqnarray}
J_{\psi,L}
:=\max\{L\}\cup\left\{1\le j \le L-1 \Big|\frac{\sum_{i=j+1}^M p_{\psi,i}^{\downarrow}}{L-j}<p_{\psi,j}^{\downarrow}\right\}.\nonumber
\end{eqnarray}
Then, there exists a suitable LOCC map to transform $\psi$ to $\eta_{\psi,L}$ \cite{Nie}, and we can obtain the following equation:
\begin{eqnarray}
&&\hspace{-1em}\max_{C:LOCC} F(C(\psi),\Phi_L)
=F(\eta_{\psi,L},\Phi_L)\nonumber\\
&&\hspace{-2em}=\sqrt{\frac{1}{L}} \displaystyle\sum_{i=1}^{J_{\psi,L}}\sqrt{p_{\psi,i}^{\downarrow}} 
+ \sqrt{\left(1-\frac{J_{\psi,L}}{L}\right){\sum_{j=J_{\psi,L}+1}^M p_{\psi,j}^{\downarrow}}}
\label{Ceq}
\end{eqnarray} 
Similarly, when we define the pure state $\zeta_{\psi,L}$ in $\mathcal{H}_{AB}$ as 
\begin{eqnarray}
\zeta_{\psi,L}= 
\sqrt{\sum_{i=1}^{L}p_{\psi,i}^{\downarrow}}^{~-1}\displaystyle\sum_{i=1}^{L}\sqrt{p_{\psi,i}^{\downarrow}}|i\rangle |i\rangle, 
\nonumber
\end{eqnarray}
 there exists a suitable LOCC map to transform $\Phi_L$ to $\zeta_{\psi,L}$, and the following holds as shown in \cite{VJN}:
\begin{eqnarray}\label{Deq}
\displaystyle\max_{D:LOCC} F(\psi,D(\Phi_L))
=F(\psi,\zeta_{\psi,L})
=\sqrt{\sum_{i=1}^{L}p_{\psi,i}^{\downarrow}}.
\end{eqnarray}
Moreover, the following equation holds for $N\le n$
\begin{eqnarray}\label{minequation}
&&\displaystyle\max_{C,D:LOCC}F(\psi^{\otimes N}, D\circ C(\psi^{\otimes n}))\nonumber\\
&&=\displaystyle\max_{D:LOCC} F(\psi^{\otimes N},D(\Phi^{\otimes m})),
\end{eqnarray}
where $C:\mathcal{S}(\mathcal{H}_{AB}^{\otimes n})\to\mathcal{S}(\mathcal{H}_{2}^{\otimes m})$ and $D:\mathcal{S}(\mathcal{H}_{2}^{\otimes m})\to\mathcal{S}(\mathcal{H}_{AB}^{\otimes N})$ run over LOCC maps.
The equation (\ref{minequation}) is obtained as follows.
At first, the right side in (\ref{minequation}) is greater than the left side since an arbitrary state in $\mathcal{S}(\mathcal{H}_{2}^{\otimes m})$ including $C(\psi^{\otimes n})$ is transformed from $\Phi^{\otimes m}$ by a suitable LOCC.
On the other hand, the right side in (\ref{minequation}) equals $F(\psi^{\otimes N},\zeta_{\psi^{\otimes N},2^{m}})$ by (\ref{Deq}), and $\zeta_{\psi^{\otimes N},2^{m}}$ can be transformed from $\psi^{\otimes n}$ via $\eta_{\psi^{\otimes n},2^{m}}$ by LOCC if $N\le n$.

Due to (\ref{minequation}), the following inequality holds. 
\begin{eqnarray}
\delta_n(N|\psi)
\ge \min_{m\in\N}  d(\psi^{\otimes n}\to\Phi^{\otimes {m}})+d(\Phi^{\otimes {m}}\to\psi^{\otimes N}).\nonumber
\end{eqnarray}
Next, we prove the converse inequality. 
Let us fix an arbitrary $m\in\N$. 
Since there exists a suitable LOCC map from $\eta_{\psi^{\otimes n},2^{m}}$ to $\zeta_{\psi^{\otimes n},2^{m}}$, we obtain 
\begin{eqnarray}\label{inequality}
\delta_n(N|\psi)
&&\le d(\eta_{\psi^{\otimes n},2^{m}},\Phi^{\otimes m})+d(\psi^{\otimes N},\zeta_{\psi^{\otimes n},2^{m}})\nonumber\\
&&= d(\psi^{\otimes n}\to\Phi^{\otimes {m}})+d(\Phi^{\otimes {m}}\to\psi^{\otimes N}).
\end{eqnarray}
We used (\ref{Ceq}) and (\ref{Deq}) to show the equality (\ref{inequality}).
 $\blacksquare$

Moreover, when the Schmidt rank of $\psi$ is $r(\psi)$, that is, $\psi$ is represented as 
\begin{eqnarray}
\psi
=\sum_{i=1}^{r(\psi)}\sqrt{p_i}|i\rangle\otimes|i\rangle,
\label{Schmidt}
\end{eqnarray}
where $|i\rangle$ is the orthonormal system,
the minimization $\min_{m\in\N}$ in (4) can be replaced by $\min_{1\le m\le N\lceil \log_2 r(\psi)\rceil}$ as follows.
From the result of Nielsen \cite{Nie},
the EPR state $\Phi^{\otimes \lceil \log_2 r(\psi)\rceil}$ can be transform to $\psi$ in (\ref{Schmidt}) without error by a suitable LOCC.
Thus, the second term in the right side of (4) equals $0$ for any $m\ge N\lceil \log_2 r(\psi)\rceil$.
Since the first term in the right side of (4) monotonically increasing with respect to $m$, the minimizer $m$ in (4) is less than $N\lceil \log_2 r(\psi)\rceil$.
This fact is important in the numerical calculation for the (generalized) MCRE in (4) and (7) and the maximal number of recoverable copies after the optimal compression process in (8).

~

\subsection{Proof of Proposition 3}
We show 
\begin{eqnarray}
&&{\lim_{n\to\infty}}d(\psi^{\otimes n} \to \Phi^{\otimes an+b\sqrt{n}})\nonumber\\
&&=\left\{
\begin{array}{ll}
0 & \mathrm{if}~a<S_{\psi} \\
G\left(\frac{b}{\sqrt{V_{\psi}}}\right) & \mathrm{if}~a=S_{\psi} \\
1 & \mathrm{if}~a>S_{\psi},
\end{array}
\right.
\label{con.asy}
\end{eqnarray}
and
\begin{eqnarray}
&&{\lim_{n\to\infty}}d(\Phi^{\otimes an+b\sqrt{n}} \to \psi^{\otimes n+b'\sqrt{n}})\nonumber\\
&&=\left\{
\begin{array}{ll}
1 & \mathrm{if}~a<S_{\psi} \\
1-G\left(\frac{b-S_{\psi}b'}{\sqrt{V_{\psi}}}\right) & \mathrm{if}~a=S_{\psi} \\
0 & \mathrm{if}~a>S_{\psi}.
\end{array}
\right.
\label{dil.asy}
\end{eqnarray}

Since the case when $a \neq S_\psi$ 
can be shown by the theorems of the direct part \cite{BBPS} and the strong converse part \cite{Hay07,HKMMW} of entanglement concentration and dilution, 
we show the case when $a = S_\psi$.
In order to show them,
we employ a function, which is similar to that used in \cite{Hay06}.
At first, we introduce a notation for this purpose.
For a Hermitian matrix $A$ and a real number $c$, 
we define the projection $\{A\le c\}$ as $\sum_{a_j\le c}P_j$, where 
the spectral decomposition of $A$ is given as
$A=\sum a_jP_j$.
Then, we introduce the following function for a sequence $\overline{\rho}=\{\rho_n\}_{n=1}^{\infty}$ of general quantum states
\begin{eqnarray}
K(b|\overline{\rho}):=
\lim_{n\to\infty}{\rm Tr}\rho_n\{\rho_n\ge 2^{-S_\psi n-b\sqrt{n}}\},
\label{general} 
\end{eqnarray}
A similar function was introduced 
in the context of the first order asymptotics for entanglement concentration \cite{Hay06}.
Then, substituting a sequence $\{\rho_{\psi}^{\otimes n+b'\sqrt{n}}\}_{n=1}^{\infty}$
into $\overline{\rho}$, we define
\begin{eqnarray}
K(b,b'|\psi) :=
K(b|\{\rho_{\psi}^{\otimes n+b'\sqrt{n}}\}_{n=1}^{\infty}),
\label{general2}
\end{eqnarray}
where $\rho_{\psi}:={\rm Tr}_B\psi$.
In the following, we simply denote $K(b,0|\psi)$ by $K(b|\psi)$.

Then, 
the central limit theorem guarantees that
\begin{eqnarray}
K(b,b'|\psi)
=G\left(\frac{b-S_{\psi}b'}{\sqrt{V_{\psi}}}\right) .
\label{K}
\end{eqnarray}
Using the relation (\ref{K}) in the case of $K(b|\psi)$,
Hayden and Winter \cite{HW03} and Harrow and Lo \cite{HL04} roughly estimated 
the accuracy of entanglement concentration and dilution based on the fidelity.
However, in order to show Proposition 3, we need 
the tight relation between $K(b|\psi)$ and the limit of the minimum 
transition errors.
For this purpose,
we show the following four inequalities for an arbitrary $\gamma>0$
\begin{eqnarray}
1-K(b+\gamma|\psi)
\le\lim_{n\to\infty}d(\Phi^{\otimes S_{\psi} n+b\sqrt{n}} \to \psi^{\otimes n}),~~\label{r-1}
\end{eqnarray}

\vspace{-2em}
\begin{eqnarray}
\lim_{n\to\infty}d(\Phi^{\otimes S_{\psi} n+b\sqrt{n}} \to \psi^{\otimes n})
\le1-K(b|\psi),\label{r-2}
\end{eqnarray}

\vspace{-2em}
\begin{eqnarray}
K(b-\gamma|\psi)
\le\lim_{n\to\infty}d(\psi^{\otimes n} \to \Phi^{\otimes S_{\psi} n+b\sqrt{n}}),\label{s-1}
\end{eqnarray}

\vspace{-2.0em}
\begin{eqnarray}
\lim_{n\to\infty}d(\psi^{\otimes n} \to \Phi^{\otimes S_{\psi} n+b\sqrt{n}})
\le K(b+\gamma|\psi).\label{s-2}
\end{eqnarray}

Then, (\ref{con.asy})
follows from (\ref{s-1}), (\ref{s-2}) and (\ref{K}),
and (\ref{dil.asy})
follows from (\ref{r-1}), (\ref{r-2}) and (\ref{K}) by changing the variable as $m=S_{\psi} n+b\sqrt{n}$ and $b=-S_{\psi}^\frac{3}{2}b'$.
Thus, it is enough to show the inequalities (\ref{r-1})-(\ref{s-2}).

At first, we prove (\ref{r-1}). 
By (\ref{Deq}), for an arbitrary state $\psi$, arbitrary positive integers ${m_n},{m'_n}$, and an arbitrary LOCC $D_n$, the inequality 
\begin{eqnarray}
F(\psi^{\otimes n},D_n(\Phi^{\otimes m_n}))^2\le\mathrm{Tr}\rho_{\psi}^{\otimes n}\{\rho_{\psi}^{\otimes n}\ge\frac{1}{2^{m'_n}}\}+\frac{2^{m_n}}{2^{m'_n}}\nonumber
\end{eqnarray}
holds. 
When ${m_n}={S_{\psi}n+b\sqrt{n}}$, ${m'_n}={S_{\psi}n+(b+\gamma)\sqrt{n}}$ in the inequality, we obtain (\ref{r-1}) by taking $\lim_{n\to\infty}$.

Next, we prove (\ref{r-2}). By (\ref{Deq}), for an arbitrary state $\psi$ and an arbitrary positive integer ${m_n}$, there is a LOCC $D_n$ which satisfies the inequality 
\begin{eqnarray}
\mathrm{Tr}\rho_{\psi}^{\otimes n}\{\rho_{\psi}^{\otimes n}\ge\frac{1}{2^{m_n}}\}\le F(\psi^{\otimes n},D_n(\Phi^{\otimes m_n}))^2.
\end{eqnarray}
When ${m_n}={S_{\psi}n+b\sqrt{n}}$ in the inequality, we obtain (\ref{r-2}) by taking the limit ${n\to\infty}$.

Next, we prove (\ref{s-1}). By Lemmas 4 and 5 in \cite{Hay06}, for an arbitrary state $\psi$, arbitrary positive integers ${m_n}\ge {m'_n}$, and an arbitrary LOCC $C_n$, the inequality 
\begin{eqnarray}
&&F(C_n(\psi^{\otimes n}),\Phi^{\otimes m_n})^2\nonumber\\
&\le&\frac{1}{2^{m_n}}\Big(\sqrt{\mathrm{Tr}\{\rho_{\psi}^{\otimes n}\ge{1}/{2^{m'_n}}\}}\sqrt{\mathrm{Tr}\rho_{\psi}^{\otimes n}\{\rho_{\psi}^{\otimes n}\ge{1}/{2^{m'_n}}\}}\nonumber\\
&&\hspace{-1.5em}+\sqrt{2^{m_n}-\mathrm{Tr}\{\rho_{\psi}^{\otimes n}\ge{1}/{2^{m'_n}}\}} \sqrt{1-\mathrm{Tr}\rho_{\psi}^{\otimes n}\{\rho_{\psi}^{\otimes n}\ge{1}/{2^{m'_n}}\}}\Big)^2 \nonumber
\end{eqnarray}
holds. 
Substituting ${S_{\psi}n+b\sqrt{n}}$ and ${S_{\psi}n+(b-\gamma)\sqrt{n}}$ into ${m_n}$ and ${m'_n}$ in the above inequality
and taking the limit ${n\to\infty}$,
we obtain (\ref{s-1}).

Finally, we prove (\ref{s-2}). 
It is enough to prove 
\begin{eqnarray}
\lim_{n\to\infty}d(\psi^{\otimes n} \to \Phi^{\otimes S_{\psi}n+b\sqrt{n}}) \le K(b+\gamma|\psi)
\end{eqnarray}
for an arbitrary positive real number $\gamma$. 
When $K(b+\gamma|\psi)$ is $1$, the inequality is obvious. 
Thus, we assume $K(b+\gamma|\psi)<1$. 
For an arbitrary positive integer ${m_n}$, 
we define the real number $x_{{m_n}}$ as a real number satisfying that
$\left\lfloor\frac{1}{x_{{m_n}}}(1-h_n(x_{{m_n}}))\right\rfloor=2^{m_n}$,
where $h_n(x):=
\mathrm{Tr}(\rho_{\psi}^{\otimes n}-x)\{\rho_{\psi}^{\otimes n}-x\ge0\}$.
By Lemma 9 and (1) in \cite{Hay06}, 
for an arbitrary state $\psi$ and an arbitrary positive integer ${m_n}$, 
there is an LOCC $C_n$ which satisfies the inequality 
\begin{eqnarray*}
1-\mathrm{Tr}\rho_{\psi}^{\otimes n}\{\rho_{\psi}^{\otimes n}\ge x_{2^{m_n}}\}
\le &F(C_n(\psi^{\otimes n}),\Phi^{\otimes m_n})^2 \\
\le &
1-d(\psi^{\otimes n} \to \Phi^{\otimes m_n}) .
\end{eqnarray*}
When we choose 
${m_n}$ as
${m_n}:={S_{\psi}n+(b+\gamma)\sqrt{n}}+\mathrm{log}(1-h_n(2^{-S_{\psi}n-(b+\gamma)\sqrt{n}}))$,
we can take $x_{{m_n}}=2^{-S_{\psi}n-(b+\gamma)\sqrt{n}}$. 
Since 
\begin{eqnarray*}
\lim_{n\to\infty}h_n(2^{-S_{\psi}n-(b+\gamma)\sqrt{n}})\le K(b+\gamma|\overline{\rho})<1,
\end{eqnarray*}
the inequality $
{S_{\psi}n+b\sqrt{n}}<{m_n}$ holds for enough large integer $n$. 
Therefore, 
\begin{eqnarray}
&&\mathrm{Tr}\rho_{\psi}^{\otimes n}\{\rho_{\psi}^{\otimes n}\ge 
2^{-S_{\psi}n-(b+\gamma)\sqrt{n}}
\}\nonumber\\
&\ge & 
d(\psi^{\otimes n} \to \Phi^{\otimes m_n})  \nonumber\\
&\ge & d(\psi^{\otimes n} \to \Phi^{\otimes {S_{\psi}n+b\sqrt{n}}}). \label{inequality10}
\end{eqnarray}
By taking the limit ${n\to\infty}$ in the inequality (\ref{inequality10}), 
we obtain (\ref{s-2}).
$\blacksquare$

We note that the existence of a lower order term $o(\sqrt{n})$ does not affect the value of $K(b,b'|\psi)$.
That is, when we define similar functions to (\ref{general}) and (\ref{general2}) as
\begin{eqnarray}
\tilde{K}(b|\overline{\rho}):=
\lim_{n\to\infty}{\rm Tr}\rho_n\{-{\rm log}\rho_n\le S_\psi n+b\sqrt{n}+o(\sqrt{n})\}\nonumber
\label{general'} 
\end{eqnarray}
and 
\begin{eqnarray}
\tilde{K}(b,b'|\psi) :=
\tilde{K}(b|\{\rho_{\psi}^{\otimes n+b'\sqrt{n}+o(\sqrt{n})}\}_{n=1}^{\infty}),\nonumber
\label{general2'}
\end{eqnarray}
then we have
\begin{eqnarray}
\tilde{K}(b,b'|\psi)
=G\left(\frac{b-S_{\psi}b'}{\sqrt{V_{\psi}}}\right)
=K(b,b'|\psi)
\label{K'}
\end{eqnarray}
because of the central limit theorem.
Thus, Proposition 3 holds as the same form even if the number of copies of $\Phi$ has a lower order term as $an+b\sqrt{n}+o(\sqrt{n})$.

~

%
\subsection{Proof of Theorem 1}
Instead of Theorem 1 itself,
we show the following more general statement for the generalized MCRE
\begin{eqnarray}
&&\lim_{n\to\infty} \delta_{n}(n+b'\sqrt{n}+o(\sqrt{n})|\psi)
\nonumber\\
&=&\left\{
\begin{array}{cl}
2G\left(\frac{S_{\psi}b'}{2\sqrt{V_{\psi}}}\right)&{\it if}~b'<0\\
1&{\it if}~b'\ge0.
\end{array}
\right.
\label{e-ineq2}
\end{eqnarray}
Since $\delta_{n}(\psi)=\delta_{n}(n|\psi)$, 
Theorem 1 is obtained from (\ref{e-ineq2}) with $b'=0$ and $o(\sqrt{n})=0$.

We set $n+b'\sqrt{n}+o(\sqrt{n})$ as $N_n$.
When $b'<0$, the inequality $N_n\le n$ holds for large $n$, and thus we can use Proposition 4.
Suppose that the number $m_n\in\N$ attains the minimum in (9) as
\begin{eqnarray}
\delta_n(N_n|\psi)
= d(\psi^{\otimes n} \to \Phi^{\otimes m_n})+d(\Phi^{\otimes m_n} \to \psi^{\otimes N_n})
\nonumber\\
\label{del2}
\end{eqnarray}
and that the asymptotic expansion of $m_n$ is $an+b\sqrt{n}+o(\sqrt{n})$.
When $a$ is not $S_{\psi}$, either the first or the second term in the right side of (\ref{del2}) goes to $1$ and the other does to $0$ as $n$ goes to $\infty$ by Proposition 3, and hence the MCRE $\delta_n(\psi)$ goes to $1$.
When $a$ is $S_{\psi}$, the first and the second term in the right side of (\ref{del2}) goes to $ G({b}/{\sqrt{V_{\psi}}})$ and $1-G({(b-S_{\psi}b')}/{\sqrt{V_{\psi}}})$, respectively, by Proposition 3.
Since a function $f(b)=G(b)+1-G(b-a)$ with $a\le0$ attains the minimum at $b=a/2$ and $G(-x)=1-G(x)$ for any $x\in\R$, 
\begin{eqnarray}
\min_{b\in\R}
G\left(\frac{b}{\sqrt{V_{\psi}}}\right)+1-G\left(\frac{b-S_{\psi}b'}{\sqrt{V_{\psi}}}\right)
=2G\left(\frac{S_{\psi}b'}{2\sqrt{V_{\psi}}}\right)
\nonumber\\
\label{2G}
\end{eqnarray}
holds.
When $b'<0$, the right side of (\ref{2G}) is less than $1$, 
and thus, we have
\begin{eqnarray}
&&\lim_{n\to\infty}\delta_n(N_n|\psi)\nonumber\\
&=&\lim_{n\to\infty} d(\psi^{\otimes n} \to \Phi^{\otimes an+b\sqrt{n}+o(\sqrt{n})})\nonumber\\
&&+d(\Phi^{\otimes an+b\sqrt{n}+o(\sqrt{n})} \to \psi^{\otimes N_n})\nonumber\\
&=&2G\left(\frac{S_{\psi}b'}{2\sqrt{V_{\psi}}}\right).
\label{2G2}
\end{eqnarray}
Thus, (\ref{e-ineq2}) was verified when $b'<0$.

When $b'\ge0$, the inequality $N_n\ge n-\epsilon\sqrt{n}$ holds for any $\epsilon>0$ and large $n$.
From (\ref{2G2}), we have
\begin{eqnarray}
\lim_{n\to\infty}\delta_n(N_n|\psi)
\ge \lim_{n\to\infty}\delta_n(n-\epsilon\sqrt{n}|\psi)
=2G\left(\frac{S_{\psi}\epsilon}{2\sqrt{V_{\psi}}}\right).\nonumber
\end{eqnarray}
By taking the limit $\epsilon\to0$ in the above inequality, we have
\begin{eqnarray}
\lim_{n\to\infty}\delta_n(N_n|\psi)
\ge1.
\label{A21}
\end{eqnarray}
On the other hand, when $C_n$ is the identity operation, the recovery error is $0$.
Then
we have 
\begin{eqnarray}
\lim_{n\to\infty}\delta_n(N_n|\psi)
\le \lim_{n\to\infty}e^{\cal C}_n(m_n,C_n|\psi)
\le1.
\label{A22}
\end{eqnarray}
From (\ref{A21}) and (\ref{A22}), we obtain (\ref{e-ineq2}) for $b'\ge0$.
$\blacksquare$

%
\subsection{Proof of Theorem 5}
We expand $N_n(\epsilon|\psi)$ as $a_{\psi,\epsilon}n+b_{\psi,\epsilon}\sqrt{n}+o(\sqrt{n})$ for $0<\epsilon<1$.
Then, in order to obtain Theorem 5, it is enough to determine the coefficients $a_{\psi,\epsilon}$ and $b_{\psi,\epsilon}$.
In general, the limit of $\delta_{n}(an+b\sqrt{n}+o(\sqrt{n})|\psi)$ for constants $a$ and $b\in\R$ is equal to $1$ for $a>1$ and is equal to $0$ for $a<1$ by (\ref{con.asy}) and (\ref{dil.asy}).
Thus, when a permissible error $\epsilon$ is between $0$ and $1$, the first order rate $a_{\psi,\epsilon}$ is $1$.
From (\ref{e-ineq2}), a suitable $b'\le0$ attains the equation
\begin{eqnarray}
\epsilon
=\lim_{n\to\infty} \delta_{n}(n+b'\sqrt{n}+o(\sqrt{n})|\psi)
=2G\left(\frac{S_{\psi}b'}{2\sqrt{V_{\psi}}}\right).\nonumber\\
\label{A23}
\end{eqnarray}
By the definition of $N_n(\epsilon|\psi)$ in (8), 
$b_{\psi,\epsilon}$ coincides with $b'$ in (\ref{A23}),
and thus 
\begin{eqnarray}
b_{\psi,\epsilon}
=-{2\sqrt{V_{\psi}}}S_{\psi}^{-1}G^{-1}\left(1-\frac{\epsilon}{2}\right).
\end{eqnarray}
Therefore, Theorem 5 was verified.


\end{document}